\documentclass[%
reprint,
showpacs,preprintnumbers,
 amsmath,amssymb,
 aps,
]{revtex4-1}

\usepackage{graphicx}

\newcommand{\myfig}[4][ht]{
\begin{figure}[#1]
\centering
\includegraphics[#2]{#3}
\caption{#4\label{#3}}
\end{figure}
}

\newcommand{\myfigwide}[4][ht]{
\begin{figure*}[#1]
\centering
\includegraphics[#2]{#3}
\caption{#4\label{#3}}
\end{figure*}
}

\graphicspath{{images/}}

\usepackage{dcolumn}
\usepackage{bm}

\begin{document}

\title{Closed periodic orbits in anomalous gravitation}

\author{Bjorn A. Vermeersch}
\email[Email: ]{vermeersch.publications@gmail.com}
\affiliation{\vspace{3mm} Grenoble, France}
\altaffiliation{The research presented here was carried out at the author's own initiative in his personal free time. No employer's resources, monetary or otherwise, were used in its creation.}

\date{\today}

\begin{abstract}
Newton famously showed that a gravitational force inversely proportional to the square of the distance, $F \sim 1/r^2$, formally explains Kepler's three laws of planetary motion. But what happens to the familiar elliptical orbits if the force were to taper off with a different spatial exponent? Here we expand generic textbook treatments by a detailed geometric characterisation of the general solution to the equation of motion for a two-body `sun/planet' system under anomalous gravitation $F \sim 1/r^{\alpha} (1 \leq \alpha < 2)$. A subset of initial conditions induce closed self-intersecting periodic orbits resembling hypotrochoids with perihelia and aphelia forming regular polygons. We provide time-resolved trajectories for a variety of exponents $\alpha$, and discuss conceptual connections of the case $\alpha = 1$ to Modified Newtonian Dynamics and galactic rotation curves.
\end{abstract}
\maketitle
\section{Introduction}
Newton's formulation of universal gravitation, and proof that it gives rise to Kepler's laws of planetary motion, stands as one of the great triumphs of analytical thinking. While relativistic corrections become necessary in certain applications, Newtonian gravity, first published 332 years ago, remains firmly in the driver's seat in space exploration, and continues to produce novel insights \cite{stewart}.
\par
Underpinning these discoveries is the central notion that any object attracts any other object with a force proportional to the product of the two masses and inversely proportional to the square of their distance: $F_{12} \sim M_1 M_2 / \| \vec{r}_1 - \vec{r}_2 \|^2$. Although the $1/r^2$ spatial dependence had previously been proposed by Hooke \cite{hookevsnewton}, Newton was the first to demonstrate mathematically that this functional form gives rise to the elliptical planetary orbits Kepler had painstakingly deduced empirically.
\par
The fact that the attractive force falls off quadratically with distance makes a lot of intuitive sense. Drawing parallels with Coulomb's electrostatic force or light emission from a star, one may argue that `gravitons' seemingly emanating from the object and exerting the gravitational pull at distance $r$ get distributed across a surface area that grows as $r^2$ in 3D space. The stunning accuracy of celestial event predictions \cite{celestialevents} and long list of highly successful space missions \cite{spacemissions} underpinned by Newtonian gravity models provide compelling empirical evidence.
\par
This, however, does not have to stop us from wondering what would happen to Kepler's neat ellipses if the gravitational force had a different spatial decay, say
\begin{equation}
F_{12} \sim \frac{1}{\| \vec{r}_1 - \vec{r}_2 \|^{\alpha}} \quad 1 \leq \alpha < 2 \label{anomalousgravity}
\end{equation}
Asking \textit{What if?}, even if only hypothetically, carries pedagogical value and often leads to thought-provoking and entertaining insights \cite{whatif}. Discussions of central force problems are available in text books such as \cite{textbook}. A later study \cite{boundedorbits} examined the generic stability and boundedness of orbits and found trajectories commonly displayed in popular texts to be physically unfeasable. Both of these cited works limited the mathematical description of the trajectories to first-order Taylor perturbations around the circular orbit.
\par
Here we expand and concretise these prior treatments by discussing the general (full-order) solution of the equation of motion. In particular, we derive for a given spatial exponent $\alpha$ the emergence and detailed geometrical properties of closed periodic orbits as function of the initial conditions. In addition, we show that a gravitational force proportional to reciprocal distance ($\alpha = 1$) has conceptual connections to Modified Newtonian Dynamics (MOND), a prominent explanation of galactic rotation curves without invoking dark matter.
\section{Equation of motion}
\subsection{Problem statement}
We wish to study the dynamics of a two-body system comprised of a primary (mass $M_1$) and satellite (mass $M_2 \ll M_1$) under a gravitational force given by
\begin{equation}
F_{\alpha} = \frac{G_{\alpha} M_1 M_2}{\| \vec{r}_1 - \vec{r}_2 \|^{\alpha}} \quad , \quad 1 \leq \alpha < 2
\end{equation}
Here $G_{\alpha}$ is a fractional gravitational constant with units m$^{\alpha+1}$kg$^{-1}$s$^{-2}$, $\vec{r}_{1,2}(t)$ denotes the time-dependent positions of the two bodies, and $\alpha$ is a characteristic exponent. The limit case $\alpha = 2$ corresponds to the familiar Newtonian one, for which $G \simeq 6.67 \times 10^{-11}\,$m$^3$kg$^{-1}$s$^{-2}$.
\par
Before exploring the equation of motion in its general form, we derive the radius $R_{\text{c}}$ and associated characteristics of the circular orbit that exists irrespective of the $\alpha$ value. These quantities will serve as normalisation values in the study of more complex orbits in dimensionless variables. In circular motion, the gravitational force provides the centripetal acceleration of the satellite:
\begin{equation}
F_{\alpha}(\| \vec{r}_1 - \vec{r}_2 \| = R_{\text{c}}) = M_2 \, \frac{v_{\text{c}}^2}{R_{\text{c}}}
\end{equation}
from which we readily obtain the linear and angular velocities in terms of the orbital radius:
\begin{equation}
v_{\text{c}} = \sqrt{\frac{G_{\alpha} M_1}{R_{\text{c}}^{\alpha-1}}} \label{vcircular}
\end{equation}
\begin{equation}
\omega_{\text{c}} = \frac{v_{\text{c}}}{R_{\text{c}}} = \sqrt{\frac{G_{\alpha} M_1}{R_{\text{c}}^{\alpha+1}}} \label{omegacircular}
\end{equation}
Invoking the conservation of angular momentum
\begin{equation}
M_2 R_{\text{c}} v_{\text{c}} = M_2 R_{\text{c}}^2 \omega_{\text{c}} = \text{constant} = L
\end{equation}
yields the orbital radius:
\begin{equation}
R_{\text{c}} = \left( \frac{L^2}{G_{\alpha} M_1 M_2^2}\right)^{1/(3-\alpha)}
\end{equation}
Finally, the orbital period is given by
\begin{equation}
T_{\text{c}} = \frac{2\pi}{\omega_{\text{c}}} = 2 \pi \sqrt \frac{R_{\text{c}}^{\alpha+1}}{G_{\alpha} M_1}
\end{equation}
\subsection{Master equation}
Central force fields confine the motion to a plane \cite{boundedorbits} for which we can introduce a polar coordinate frame $(r,\varphi)$ with the primary at the center $r=0$. Applying Newton's second law to the satellite in anomalous gravitation yields
\begin{equation}
\frac{\mathrm{d}^2 r}{\mathrm{d} t^2} - r \left( \frac{\mathrm{d}\varphi}{\mathrm{d}t} \right)^2 = - \frac{G_{\alpha} M_1}{r^{\alpha}}
\end{equation}
We now invoke conservation of angular momentum
\begin{equation}
M_2 \, r^2(\varphi) \, \frac{\mathrm{d}\varphi}{\mathrm{d}t} = M_2 \, r^2(\varphi) \, \omega(\varphi) = \text{constant} = L \label{angmom}
\end{equation}
and revert to dimensionless coordinates
\begin{equation}
\bar{r} \equiv \frac{r}{R_{\text{c}}}
\end{equation}
where $R_{\text{c}}$ is the radius of the circular orbit with same angular moment $L$ discussed previously. A subsequent change of variable
\begin{equation}
u(\varphi) \equiv \frac{1}{\bar{r}(\varphi)} = \frac{R_{\text{c}}}{r(\varphi)}
\end{equation}
finally produces
\begin{equation}
\frac{\mathrm{d}^2 u}{\mathrm{d} \varphi^2} + u(\varphi) = \frac{1}{u(\varphi)^{2-\alpha}} \label{mastereq}
\end{equation}
Without loss of generality, we can restrict ourselves to initial conditions of the form
\begin{equation}
u(0) = u_0 \geq 1 \quad , \quad u'(0) \equiv \frac{\mathrm{d}u}{\mathrm{d}\varphi} \biggr |_{\varphi = 0} = 0 \label{masteric}
\end{equation}
because, as shown below, the general solution for $u$ is periodic in $\varphi$. One can therefore always perform a rotation of the coordinate axes that places a maximum of $u$ at $\varphi = 0$, i.e. the orbit has a perihelion on the $x$ axis.
\subsection{Angular and linear velocities}
Conservation of angular momentum (\ref{angmom}) readily yields
\begin{equation}
\omega(\varphi) = \frac{L}{M_2 \, r^2(\varphi)} = \frac{L}{M_2 R_{\text{c}}^2} \, \frac{1}{\bar{r}^2(\varphi)} = \omega_{\text{c}} \, u^2(\varphi) \label{angularvelocity}
\end{equation}
Points with extremal angular velocity obey
\begin{equation}
\frac{\mathrm{d}\omega}{\mathrm{d}\varphi} = 0 \quad \leftrightarrow \quad 2 \, u(\varphi) \, \frac{\mathrm{d}u}{\mathrm{d}\varphi} = 0
\end{equation}
As the first factor remains strictly positive for bounded orbits, the maximal and minimal angular velocities are achieved in the perihelia and aphelia respectively.
\par
The linear velocity of the satellite is given by
\begin{equation}
v(\varphi) \equiv \left\| \frac{\mathrm{d}\vec{r}}{\mathrm{d}t} \right\| = \omega(\varphi) \sqrt{ \left( \frac{\mathrm{d}r}{\mathrm{d}\varphi}\right)^2 + r^2(\varphi)}
\end{equation}
Using (\ref{angularvelocity}) and $v_{\text{c}} = \omega_{\text{c}} R_{\text{c}}$, the linear velocity can be expressed directly in terms of the inital value problem solution $u(\varphi) = R_{\text{c}}/r(\varphi)$ as
\begin{equation}
v(\varphi) = v_{\text{c}} \sqrt{ \left( \frac{\mathrm{d}u}{\mathrm{d}\varphi}\right)^2 + u^2(\varphi)} \label{velocity}
\end{equation}
The velocity reaches an extremal value when
\begin{equation}
\frac{\mathrm{d}v}{\mathrm{d}\varphi} = 0 \quad \leftrightarrow \quad \frac{\mathrm{d}u}{\mathrm{d}\varphi} \cdot \left[ \frac{\mathrm{d}^2 u}{\mathrm{d}\varphi^2} + u(\varphi) \right] = 0
\end{equation}
Per (\ref{mastereq}), the factor between brackets equals $u^{\alpha-2}(\varphi)$, which remains strictly positive for bounded orbits. Just like the angular counterpart, the linear velocity is largest at perihelion and smallest at aphelion:
\begin{equation}
v_{\text{max}} = v_{\text{c}} \, u_0 \quad , \quad v_{\text{min}} = v_{\text{c}} \, u_{\text{min}} \label{vextremal}
\end{equation}
\subsection{Orbital position as function of time}
Rewriting (\ref{angularvelocity}) yields
\begin{equation}
\mathrm{d}\varphi = \omega_{\text{c}} \, u^2 (\varphi) \, \mathrm{d}t \quad \Rightarrow \quad t(\varphi) = \frac{T_{\text{c}}}{2\pi} \int \limits_{0}^{\varphi} \frac{\mathrm{d}\varphi'}{u^2(\varphi')} \label{orbitaltime}
\end{equation}
This offers an explicit mapping between the polar angle of points along the orbit and the time elapsed to reach them. Using the relation in the reverse direction provides the position of the satellite at any given time.
\subsection{Orbital bounds}
Previous stability analysis has shown that orbits generated by the spatial exponents considered here ($1 \leq \alpha \leq 2 $) `can be bounded or unbounded depending on the total (satellite) energy' \cite{boundedorbits}. Here we establish explicit orbital bounds for a given $\alpha$ as a function of $u_0$. As mentioned previously, the perihelion is prescribed by the initial conditions (\ref{masteric}) themselves:
\begin{equation}
r_{\text{min}} = \frac{R_{\text{c}}}{u_{\text{max}}} = \frac{R_{\text{c}}}{u_0}
\end{equation}
The aphelion distance $r_{\text{max}} = R_{\text{c}}/u_{\text{min}}$ can be derived from the conservation of total energy. Indeed, the amount of kinetic energy the satellite loses upon moving from perihelion to aphelion must precisely equal the work performed against the gravitational field to increase the distance to the primary from $r_{\text{min}}$ to $r_{\text{max}}$:
\begin{equation}
\frac{M_2}{2} \left( v_{\text{max}}^2 - v_{\text{min}}^2 \right) = \int \limits_{r_{\text{min}}}^{r_{\text{max}}} \frac{G_{\alpha} M_1 M_2}{r^{\alpha}} \, \mathrm{d}r
\end{equation}
Invoking (\ref{vcircular}) and (\ref{vextremal}) leads to
\begin{equation}
\begin{cases}
u_{\text{min}} = 2 - u_0  & \alpha = 2\\
u_0^2 - u_{\text{min}}^2 = \frac{2}{\alpha - 1} \left( u_0^{\alpha - 1} - u_{\text{min}}^{\alpha - 1} \right) & 1 < \alpha < 2\\
u_0^2 - u_{\text{min}}^2 = 2 \, (\ln u_0 - \ln u_{\text{min}}) & \alpha = 1 \\
\end{cases}
\label{umin_analytic}
\end{equation}
Bounded orbits require strictly positive $u_{\text{min}}$ ($r_{\text{max}}$ should remain finite), leading to the criterion
\begin{equation}
u_0 < \left( \frac{2}{\alpha - 1} \right)^{1/(3-\alpha)} \quad 1 < \alpha \leq 2
\end{equation}
The remainder of the paper will focus on initial conditions $1 \leq u_0 < 2$, which produce bounded orbits for all exponents $1 \leq \alpha \leq 2$.
\section{Bounded orbits}
\subsection{Circular orbits} \label{circular}
The master equation (\ref{mastereq}) supports a constant solution:
\begin{equation}
u(\varphi) = 1 \quad \leftrightarrow \quad r(\varphi) = R_{\text{c}}
\end{equation}
which corresponds to the circular orbit discussed earlier. One easily verifies that inserting $u(\varphi) = 1$ into (\ref{angularvelocity}), (\ref{velocity}) and (\ref{orbitaltime}) respectively recovers $\omega(\varphi) = \omega_{\text{c}}$, $v(\varphi) = v_{\text{c}}$ and $t(\varphi = 2\pi) = T_{\text{c}}$ as appropriate.
\subsection{Near-circular orbits} \label{subsec:nearcircular}
Let us consider initial conditions of type
\begin{equation}
u_0 = 1 + e \quad \text{with} \quad 0 < e \ll 1
\end{equation}
These lead to previously published \cite{boundedorbits,textbook} solutions
\begin{equation}
u(\varphi) \simeq 1 + e \, \cos (\beta_{e} \varphi) \quad , \quad \beta_{e} = \sqrt{3-\alpha} \label{nearcircular}
\end{equation}
as can be verified by substituting this form into the master equation, performing a series expansion $1/u^{2-\alpha} \simeq 1 - (2-\alpha) e \cos (\beta_{e} \varphi)$, and equating the constant and first harmonic terms on the left and right hand sides.
\par
In the Newtonian limit $\alpha = 2$ (yielding $\beta_{e} = 1$) the solution becomes exact regardless the magnitude of $e$, and corresponds to the familiar elliptical ($0 < e < 1$) or hyperbolic ($e>1$) orbits with the primary as focus.
\par
The trajectories generated for $0 < e \ll 1$ by other exponents $1 \leq \alpha < 2$ remain within close vicinity of the circular orbit since $1 - e \lesssim \bar{r}(\varphi) \lesssim 1 + e$. Interestingly, the orbits approximately follow hypotrochoids traced out by a circle with radius $R_{\text{c}}/(\beta_e-1)$ inside a larger circle with radius $\beta_e R_{\text{c}}/(\beta_e-1)$ (see Appendix \ref{app:hypotrochoids}).
\subsection{General solutions} \label{subsec:generalsolution}
The approximation (\ref{nearcircular}) becomes progressively less accurate when the initial point moves inwards, as higher-order terms ought to be considered in the series expansion of $1/u^{2-\alpha}$ when $e$ can no longer be considered as very small. However, since both $\mathrm{d}^2/\mathrm{d}\varphi^2$ and fractional power operators conserve the periodicity of the input function, the master equation clearly supports periodic $u(\varphi)$, and we can postulate a general solution of the form
\begin{equation}
u(\varphi) = \sum \limits_{n=0}^{\infty} A_n \cos (n \beta \varphi)  
\end{equation}
An approximate realisation truncated to the $N$ lowest harmonics can conceptually be constructed as follows:
\begin{enumerate}
\item{Perform a series expansion $1/(1+\chi)^b \simeq 1 - b \chi + \frac{1}{2} b (b+1) \chi^2 - \cdots$ of $1/u^{2-\alpha}$ up to $N$-th order.}
\item{Rewrite resulting powers $\cos^m(n \beta \varphi)$ as linear combinations of harmonics via de Moivre's formula.}
\item{Equate the coefficients of the static term and lowest $N$ harmonics on either side.}
\item{Enforce the initial condition: $\sum \limits_{n=0}^{N} A_n = u_0$.}
\end{enumerate}
Steps 3 and 4 together produce a set of $N+2$ polynomial equations in $N+2$ unknowns, being $N+1$ spectral coefficients $\left\{A_n\right\}_{n=0}^{N}$ and the periodicity factor $\beta$. Explicit formulae and numeric examples for $N=2$ are listed in Appendix \ref{app:twoharmonics}. The procedure quickly becomes cumbersome for practical use as $N$ increases. Accurate solutions can instead be obtained by direct numerical integration of the initial value problem, as will be done in Sec. \ref{sec:results}.
\par
The presence of higher-order harmonics causes orbits to quantitatively deviate more and more from true hypotrochoids as $u_0-1$ becomes progressively larger, though a strong qualitative resemblence remains.
\subsection{Closed periodic orbits}
An interesting subset comprises solutions having rational $\beta = p/q$ ($p$ and $q$ coprime integers). These yield closed orbits that are traversed cyclically every $q$ revolutions, with extremal points forming 2 regular $p$-gons: 
\begin{eqnarray}
\text{perihelia} & : & \varphi = \frac{mq}{p} \, 2\pi \quad (m \in \mathbb{N}) \\
\text{aphelia} & : & \varphi = \frac{(m + \frac{1}{2}) q}{p} \, 2\pi \quad (m \in \mathbb{N})
\end{eqnarray}
Each vertex is visited once per orbital cycle, in a sequence that moves $p-q$ point(s) clockwise along the polygons.
\section{Results}\label{sec:results}
All results presented here were obtained by numerically solving the initial value problem (\ref{mastereq})--(\ref{masteric}) through the Dormand-Prince method with absolute and relative error tolerances set to $10^{-6}$. Software implementations of this 4th/5th order explicit Runge-Kutta scheme are available across multiple applications and platforms, including the \texttt{ode45} solver in Octave \cite{ode45}.
\par
Each orbit is plotted in dimensionless cartesian coordinates $(x/R_{\text{c}}, y/R_{\text{c}})$ i.e. normalised to the radius of the circular orbit having the same angular momentum (displayed for visual reference in grey shading).
\subsection{Orbital bounds of general solution}
Figure \ref{fig1_boundedness} shows the dimensionless aphelion distance $r_{\text{max}}/R_{\text{c}}$ as a function of the initial condition $u_0$ for characteristic exponents $\alpha = 2, 1.8, \ldots , 1$. For a given $u_0$ (corresponding to the same initial normalised kinetic energy $v_{\text{max}}^2 / v_{\text{c}}^2$ for all $\alpha$), the orbit is systematically more tightly bound to the primary as $\alpha$ becomes smaller, because gravitational fields that decay less sharply in space require more work to increase orbital distance.
\myfig[!htb]{width=0.4\textwidth}{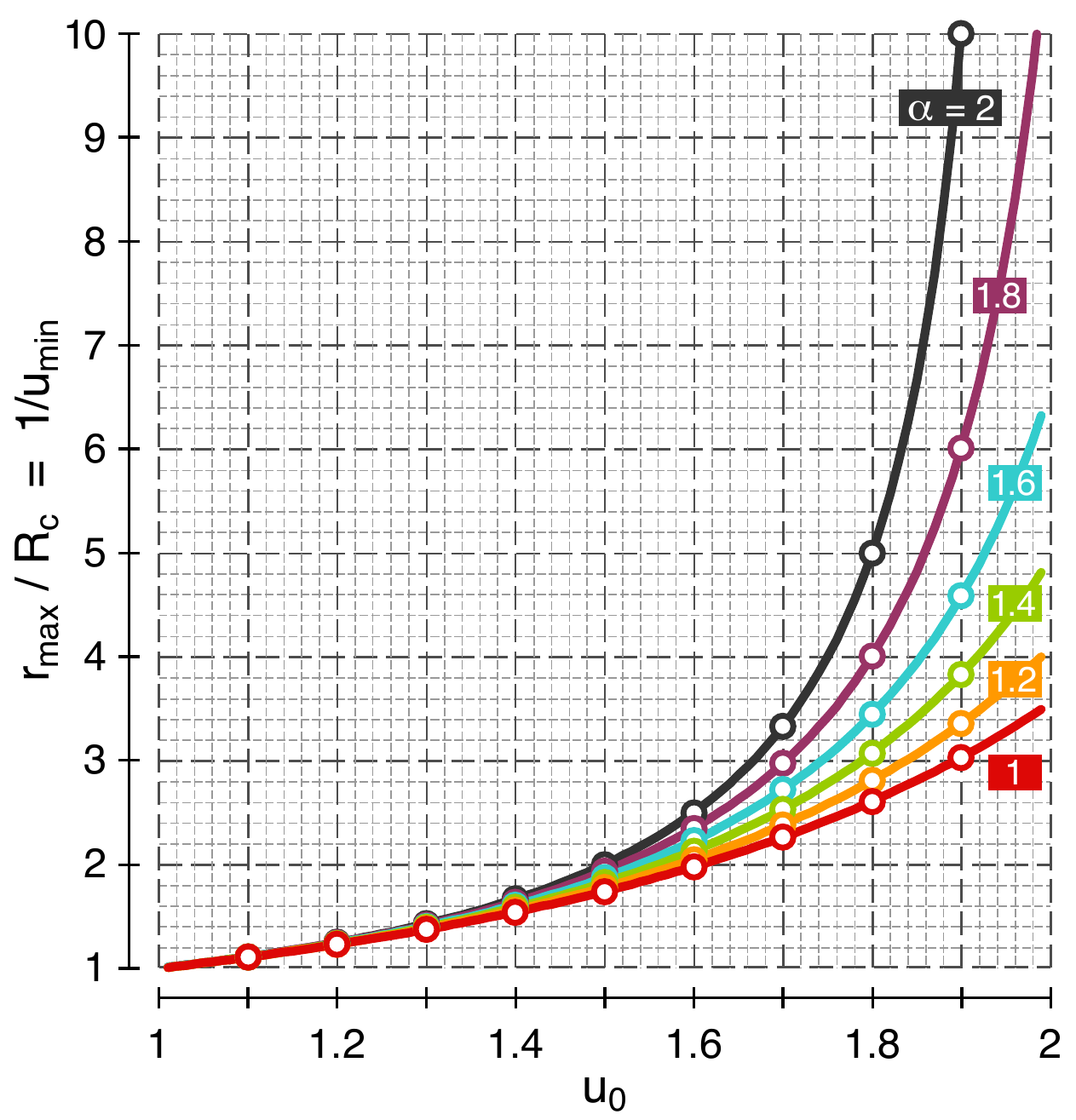}{Aphelion distance normalised to radius $R_c$ of the circular orbit. Values observed in numerically computed orbits (open circles) agree perfectly with the analytic result (\ref{umin_analytic}) derived from conservation of energy (lines).}
\subsection{Periodicity of general solution}
Figure \ref{fig2_periodicity} shows the periodicity factor $\beta$ as a function of the initial condition $u_0$ for characteristic exponents $\alpha = 2, 1.9, \ldots , 1$. Results were obtained to 4 significant digits by determining the period $2\pi/\beta$ of numerical solutions $u(\varphi)$ computed with step size $\Delta \varphi = 10^{-4}\,$rad. The $\beta(u_0)$ dependence systematically intensifies towards larger $u_0$ values (more eccentric orbits) and more anomalous exponents (lower $\alpha$ values). We remind that the Newtionian case $\alpha = 2$ is characterised by $\beta \equiv 1$ exactly for all initial conditions.
\myfig[!htb]{width=0.48\textwidth}{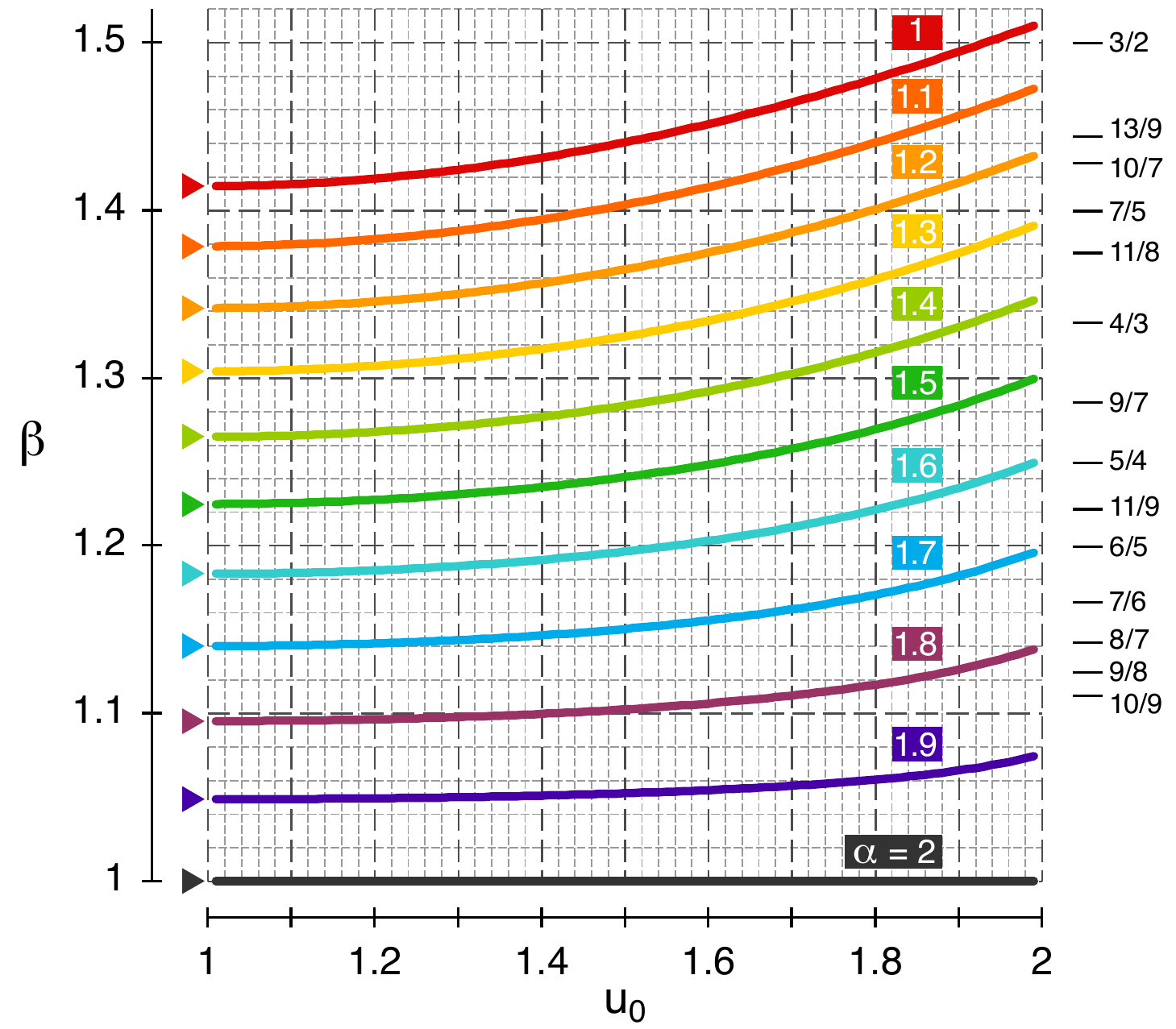}{Periodicity factor $\beta$ for the general solution $u(\varphi) = \sum A_n \cos(n \beta \varphi)$ of the initial value problem (\ref{mastereq})--(\ref{masteric}). The triangles on the left indicate $\beta_e = \sqrt{3-\alpha}$ derived for near-circular orbits ($u_0 \simeq 1$). The lines on the right mark exemplary rational values $\beta = p/q$ inducing closed periodic orbits.}
\subsection{Geometric features}
Figure \ref{fig3_hypotrochoids} illustrates the close connection between the orbital shapes induced by anomalous gravitation laws and hypotrochoids. The perihelia and aphelia of closed periodic orbits (rational periodicity factor $\beta$) moreover form the vertices of regular (star) polygons (Fig. \ref{fig4_polygons}).
\myfigwide[!htb]{width=\textwidth}{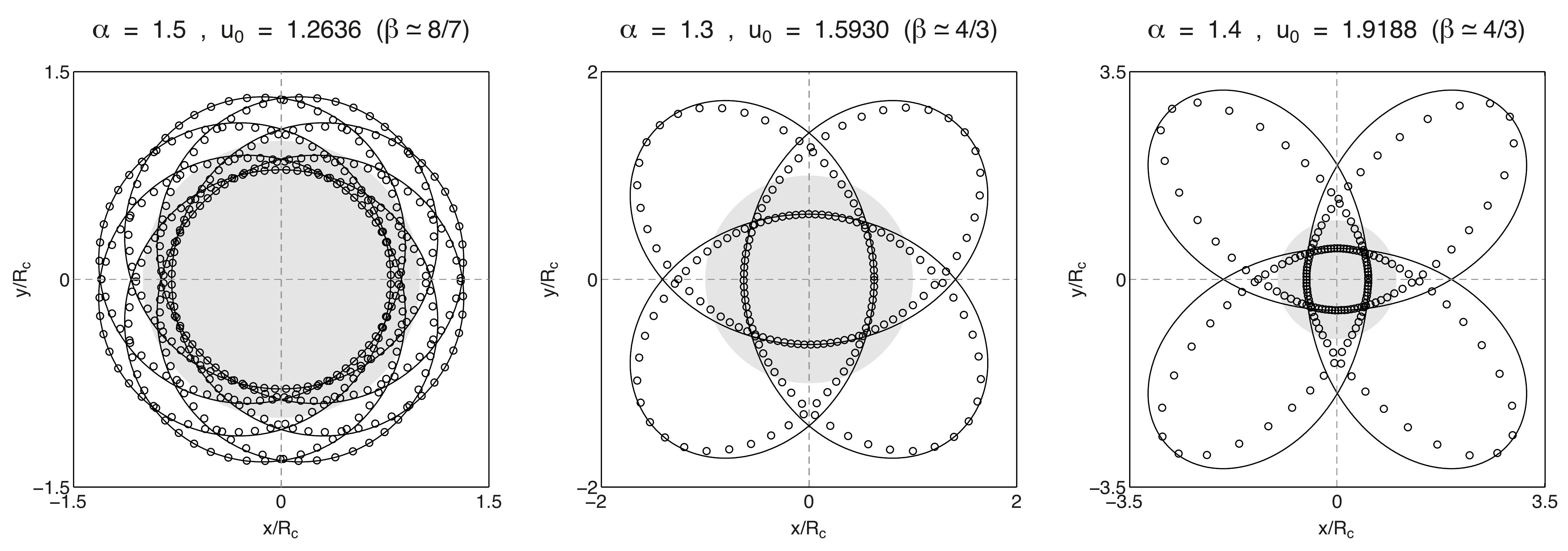}{Computed orbits (open circles) closely resemble hypotrochoids (solid lines). The quantitative agreement is best for moderate initial conditions $(u_0 \lesssim 1.3)$ which induce orbits in the vicinity of the circular one (grey shading). More eccentric orbits (larger $u_0$) have sharper lobes than their hypotrochoidal counterparts but a clear qualitative correspondence remains.}
\myfigwide[!htb]{width=\textwidth}{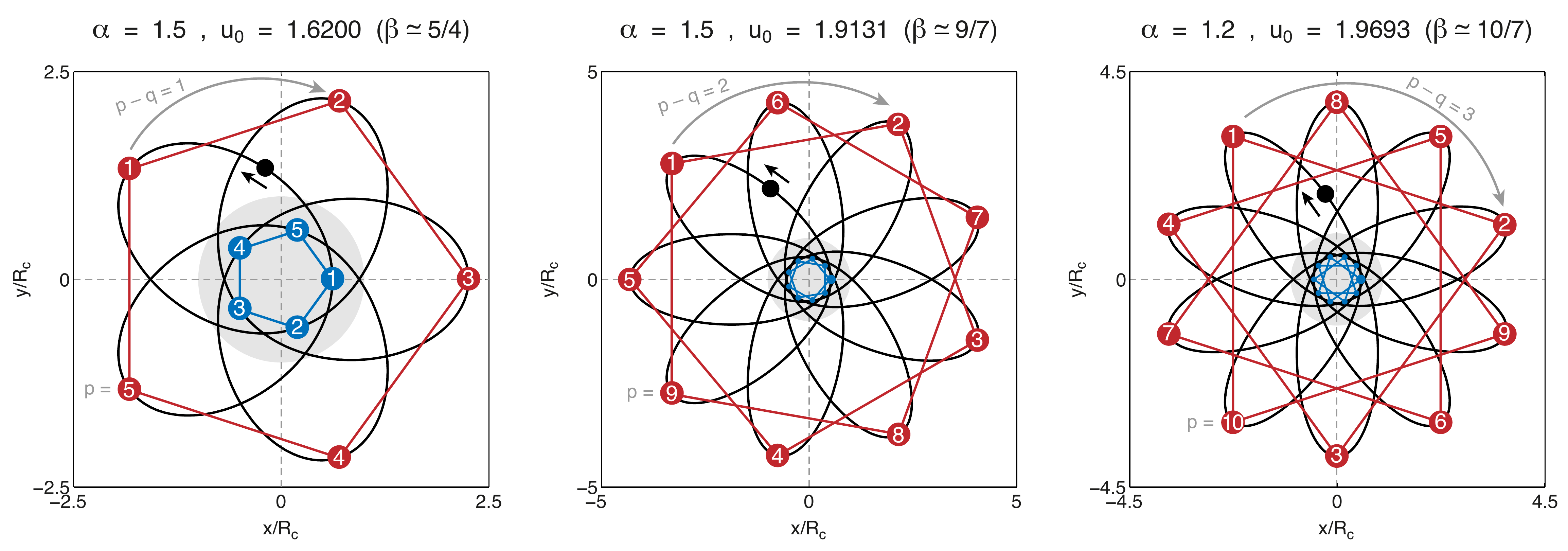}{Rational periodicity factors $\beta = p/q$ with $p$ and $q$ coprime induce closed periodic orbits (black lines) whose perihelia (blue points) and aphelia (red points) form regular $p$-gons. A satellite revolving around the primary counter-clockwise visits each of the vertices once per orbital cycle in a sequence that moves $p-q$ point(s) clockwise along the polygons.}
\subsection{Time-resolved closed periodic orbits}
Figures \ref{fig5_orbits1} and \ref{fig6_orbits2} show various examples of closed periodic orbits across a range of characteristic exponents $\alpha$. Each orbit is plotted in 3 color-coded renderings that respectively show dimensionless angular velocity, linear velocity, and time (normalised to circular orbit counterparts $\omega_{\text{c}}$, $v_{\text{c}}$ and orbital period $T_{\text{c}}$ respectively).
\myfigwide[!htb]{width=0.98\textwidth}{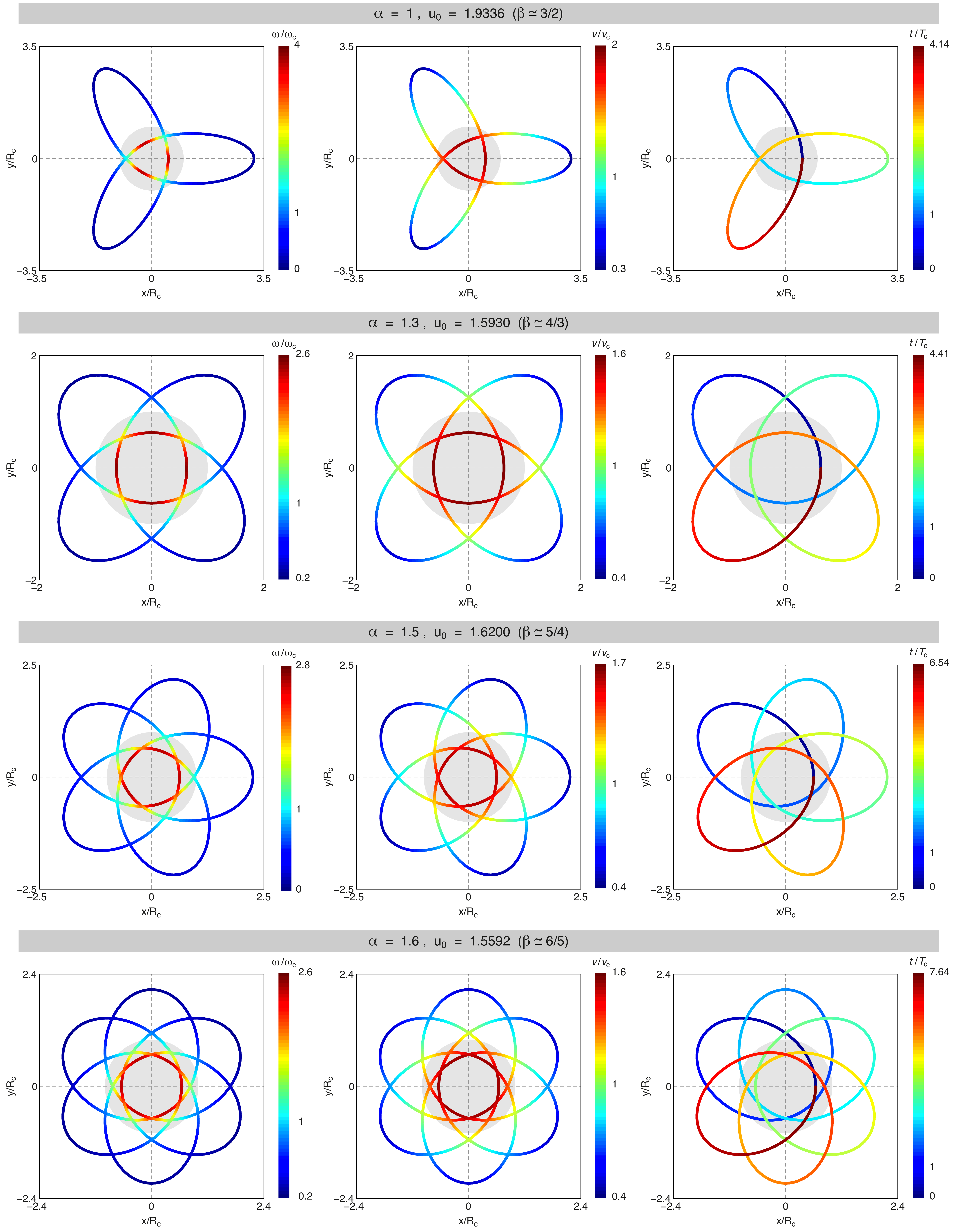}{Examples of closed periodic orbits. The color coding shows angular velocity (left column), linear velocity (middle column), and time (right column), each normalised to counterparts of the circular orbit $r = R_{\text{c}}$ (grey shading). Time-stepped video animations of each of these orbits are available as supplementary media files.}
\myfigwide[!htb]{width=0.98\textwidth}{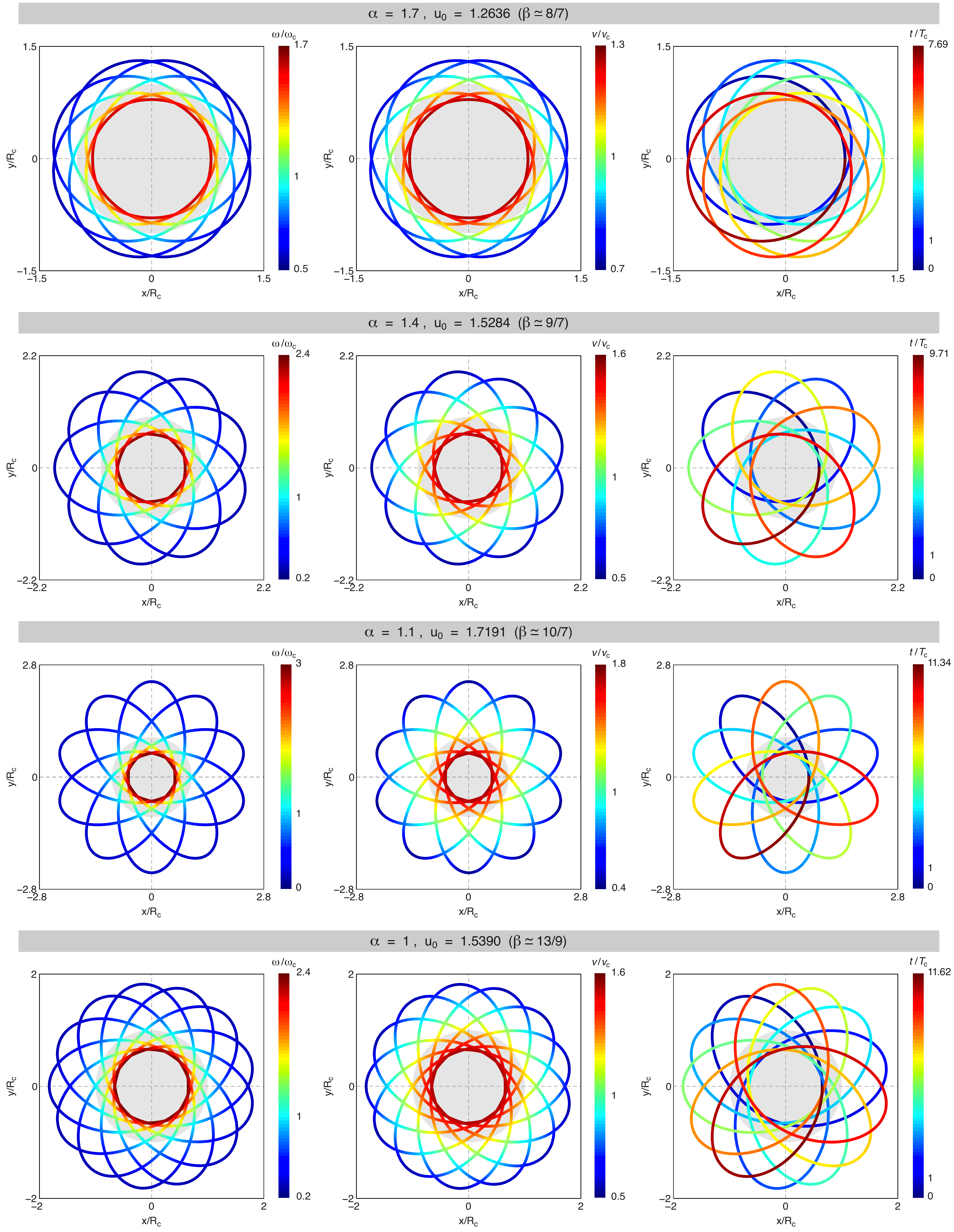}{Examples of closed periodic orbits. The color coding shows angular velocity (left column), linear velocity (middle column), and time (right column), each normalised to counterparts of the circular orbit $r = R_{\text{c}}$ (grey shading). Time-stepped video animations of each of these orbits are available as supplementary media files.}
\section{Connection with MOND: \newline the case $\alpha = 1$}
Under Newtonian gravity the velocity of a star in circular orbit around a galactic core reads $v(R) = \sqrt{G M_R / R}$ where $M_R$ denotes the total galactic mass contained within the star's orbital radius $R$. However, empirical rotation curves $v(R)$ do not decay as the expected $1/\sqrt{R}$ but become flat instead: $v(R) \sim \text{constant}$ \cite{rotationcurves}. Two main explanations for this this discrepancy between predicted and observed behaviour have emerged \cite{rotationcurves}. 
\par
On the one hand, the `dark matter' hypothesis postulates that galaxies possess haloes of unknown gravitationally interacting substance not detectable by current means. This effectively adds mass to the $M_R$ value inferred from regular observable matter, which counteracts the $1/\sqrt{R}$ velocity drop-off. Despite considerable efforts by cosmologists and particle phycisists, dark matter to date has never been observed directly, and remains a hypothetical substance unknown to physics \cite{stewart}.
\par
Modified Newtonian dynamics (MOND), on the other hand, maintains nominal $M_R$ values but instead posits that Newton's second law breaks down over galactic scales, with forces becoming proportional to the square of the acceleration when the latter is very small \cite{mond}. Formally, MOND postulates that
\begin{equation}
F = m \, \mu \hspace{-0.2em} \left( \frac{a}{a_0} \right) a \label{MONDmaster}
\end{equation}
Here $\mu(x)$ is an interpolation function with asympotes $\mu(x \ll 1) \sim x$ and $\mu(x \gg 1) \sim 1$, e.g. of the form
\begin{equation}
\mu(x) = \frac{1}{\left[ 1+(1/x)^{\gamma} \right]^{1/\gamma}} \quad , \quad \gamma > 0
\end{equation}
and $a_0 \simeq 1.2 \times 10^{-10}\,$m/s$^2$ \cite{mond} is an empirical fitting constant acting as a characteristic acceleration below which non-Newtonian behaviour becomes important. Applying (\ref{MONDmaster}) with $\gamma = 1$ for simplicity to a circular orbit with radius $R$ under Newtonian gravity shows that
\begin{eqnarray}
& & \frac{G M_R}{R^2} = \frac{a^2}{a + a_0} = \frac{v^4/R^2}{v^2/R + a_0} \\
& & \Leftrightarrow \begin{cases}
a \gg a_0 \text{ (Newtonian)} : & v = \sqrt{G M_R/R} \\[2mm]
a \ll a_0 \text{ (deep MOND)} : & v = \sqrt[4]{G M_R a_0}
\end{cases}
\label{vMOND}
\end{eqnarray}
where the latter is indeed independent of $R$.
\par
Alternatively, one could maintain the second law but instead modify the gravitational force, say
\begin{equation}
F_{12}(r) = \frac{G M_1 M_2}{r^2} \left( 1 + \frac{r}{\mathcal{R}_0} \right)
\end{equation}
At small distances $r \ll \mathcal{R}_0$ the force is Newtonian, while large distances $r \gg \mathcal{R}_0$ experience anomalous gravitation of type (\ref{anomalousgravity}) with exponent $\alpha = 1$ and fractional constant $G_{\alpha} = G/\mathcal{R}_0$. Applying Newton's unmodified second law to a circular orbit with radius $R$ now yields
\begin{equation}
\begin{cases}
R \ll \mathcal{R}_0 \text{ ($\alpha = 2$)} : & v = \sqrt{G M_R/R} \\[2mm]
R \gg \mathcal{R}_0 \text{ ($\alpha = 1$)} : & v = \sqrt{G M_R/\mathcal{R}_0} \end{cases}
\end{equation}
where once again the latter is independent of orbital radius. Comparison with (\ref{vMOND}) shows that
\begin{equation}
\mathcal{R}_0 = \sqrt{\frac{G M_R}{a_0}}
\end{equation}
For galactic bulge masses typically on the order of $10^{10}$ Suns ($M_R \simeq 2 \times 10^{40}\,$kg) we find
\begin{equation}
\mathcal{R}_0 \simeq 3.4 \, \text{kiloparsecs} \simeq 11,000 \, \text{light years}
\end{equation}
which is indeed comparable to the orbital distances at which rotation curves begin to flatten out \cite{rotationcurves}.
\section{Conclusions}
We deduced detailed characteristics of bounded orbits under anomalous gravitational forces $F \sim 1/r^{\alpha}$ with spatial exponents $1\leq \alpha < 2$. The familiar Keplerian ellipses make way for hypotrochoid-like self-intersecting trajectories. Despite their convoluted appearance for most initial conditions, the orbits are governed by a general solution that is, just like the Newtonian one, periodic in polar angle. A subset of cases exhibit a `mathematical resonance' in which this periodic solution is traversed an integer $p$ times for every $q$ revolutions. These lead to fully closed periodic orbits with the primary as centroid and perihelia and aphelia forming regular $p$-gons.
\par
Overall, the key differences between anomalous and Newtonian gravity could be summarised in terms of a heightened sensitivity to initial conditions for the former. Indeed, moving the launch site across the bounded range will drastically alter the orbit's geometric appearance (whether it is closed or not, and if yes how many `lobes' it contains), while the Keplerian counterpart remains elliptical but simply with a different eccentricity.
\par
Some features do remain preserved, as well: all spatial exponents support circular orbits, albeit with distinct radii and orbital velocities. The latter becoming independent of the radius for a gravitational force inversely proportional to distance, the anomalous case $\alpha = 1$ is compatible with galactic rotation curves and conceptually linked to Modified Newtonian Dynamics.
%
\clearpage
\appendix
\section{Link with hypotrochoids} \label{app:hypotrochoids}
Let us consider a circle with radius $R_1$ rolling inside a larger circle with radius $R_2 = \beta R_1$. (Fig. \ref{figA1_hypotrochoid}).
\myfig[!htb]{width=0.35\textwidth}{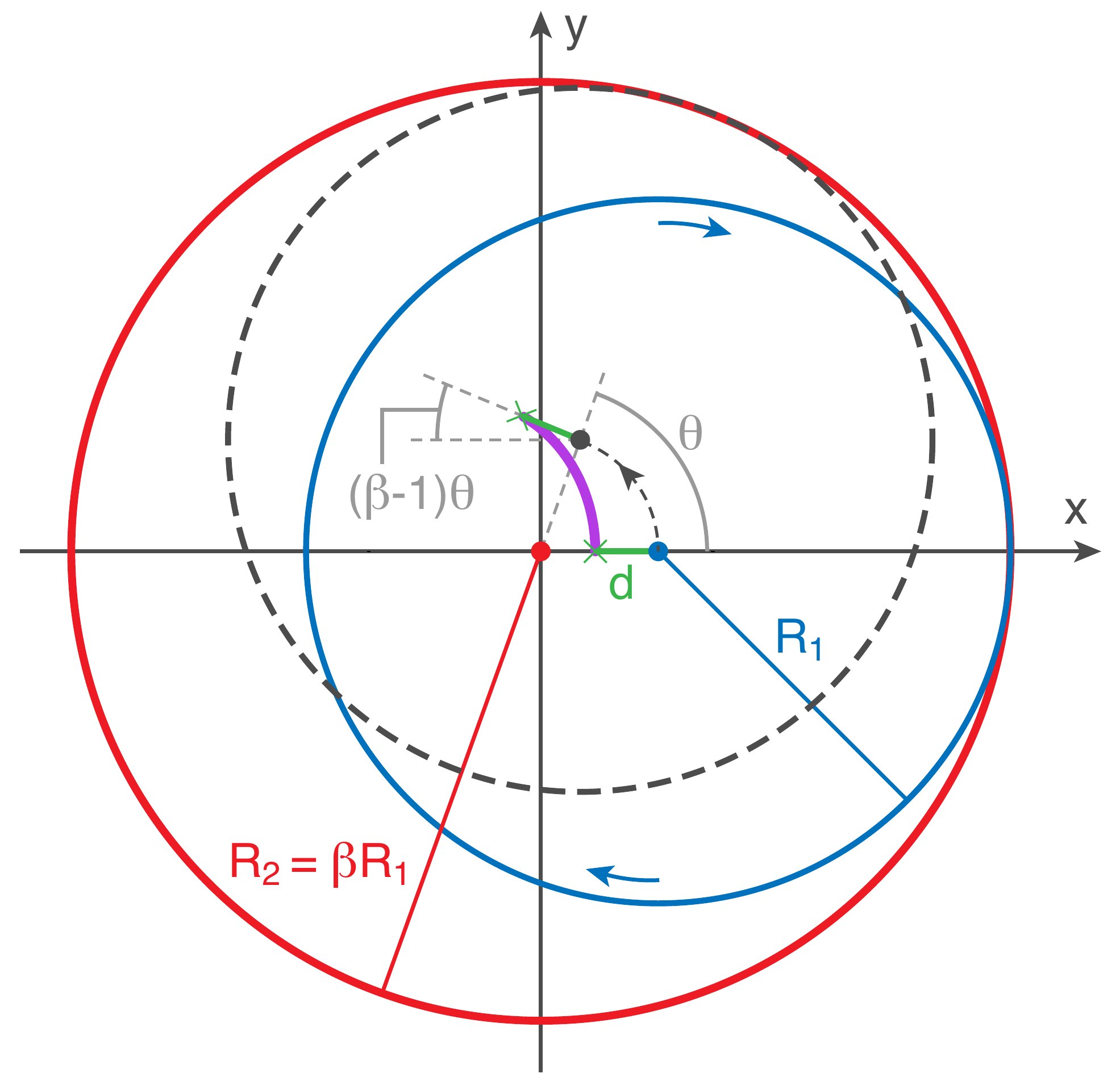}{Construction of hypotrochoidal curve (purple) traced out by a point (marked $\times$) a distance $d$ from the center of a smaller circle with radius $R_1$ (blue) rolling without slipping inside a larger circle with radius $R_2 = \beta R_1$ (red). The depicted example produces the curve shown in Fig. \ref{fig3_hypotrochoids}b.}
\par
A point on the small circle at distance $d$ from its center and initially between the two circle centers traces a curve parametrised by
\begin{equation}
\begin{cases}
x = (\beta-1) R_1 \cos \theta - d \cos [(\beta-1) \theta] \\
y = (\beta-1) R_1 \sin \theta + d \sin [(\beta-1) \theta]
\end{cases}
\end{equation}
where $\theta$ is the angle marking the position of the center of the small circle. In polar coordinates we have
\begin{equation}
\begin{cases}
r^2(\varphi) = (\beta-1)^2 R_1^2 + d^2 - 2(\beta-1) R_1 d \cos [\beta \theta(\varphi)] \\[2mm]
\tan \varphi = \frac{(\beta-1) R_1 \sin \theta + d \sin [(\beta-1) \theta]}{(\beta-1) R_1 \cos \theta - d \cos [(\beta-1) \theta]}
\end{cases}
\end{equation}
Setting $R_1 = 1/(\beta - 1)$ and assuming $d \ll 1$ yields
\begin{equation}
\begin{cases}
r^2(\varphi) \simeq 1 - 2 d \cos (\beta \theta) \\[2mm]
\tan \varphi \simeq \frac{\sin \theta}{\cos \theta} = \tan \theta \quad \leftrightarrow \quad \varphi \simeq \theta
\end{cases}
\label{hypocurve}
\end{equation}
Meanwhile, solutions $u(\varphi) \simeq 1 + e \cos (\beta \varphi)$ with $e \ll 1$ of the initial value problem (\ref{mastereq})--(\ref{masteric}), produce orbits
\begin{equation}
\bar{r}^2(\varphi) \equiv \frac{1}{u(\varphi)^2} \simeq 1 - 2 e \cos(\beta \varphi)
\end{equation}
which describes the same curve as (\ref{hypocurve}) if setting $d=e$.
\par
For a given general solution $u(\varphi)$ (not necessarily inducing a near-circular orbit), the hypotrochoid having the same periodicity factor $\beta$ and same extremal radii $\bar{r}_{\text{min}} = 1/u_0$ , $\bar{r}_{\text{max}} = 1/u_{\text{min}} = 1/u(\pi/\beta)$ is generated by
\begin{equation}
R_1 = \frac{R_2}{\beta} = \frac{\bar{r}_{\text{max}} + \bar{r}_{\text{min}}}{2(\beta -1)} \quad , \quad d = \frac{\bar{r}_{\text{max}} - \bar{r}_{\text{min}}}{2}
\end{equation}
\section{Explicit formulae for general solution truncated to 2 harmonics}\label{app:twoharmonics}
We seek an approximate solution of the initial value problem (\ref{mastereq})--(\ref{masteric}) of the form
\begin{equation}
u(\varphi) \simeq A_0 + A_1 \cos (\beta \varphi) + A_2 \cos (2 \beta \varphi)
\end{equation}
Executing the procedure outlined in Sec. \ref{subsec:generalsolution} produces the following set of equations:
\begin{equation}
\begin{cases}
A_0^{3-\alpha} = 1 + \frac{1}{4}(2-\alpha)(3-\alpha) \left[ \frac{A_1^2}{A_0^2} + \frac{A_2^2}{A_0^2} \right] \\
-(\beta^2-1) A_1 A_0^{2-\alpha} = -(2-\alpha) \frac{A_1}{A_0} \left[ 1 - \frac{1}{2} (3-\alpha) \frac{A_2}{A_0} \right] \\
-(4\beta^2-1) A_2 A_0^{2-\alpha} = -(2-\alpha) \frac{A_1}{A_0} \left[ \frac{A_2}{A_1} - \frac{1}{4} (3-\alpha) \frac{A_1}{A_0} \right] \\
A_0 + A_1 + A_2 = u_0
\end{cases}
\label{eqs2harmonics}
\end{equation}
These can be solved numerically for given pairs of gravitational exponent $\alpha$ and initial value $u_0$ to obtain the periodicity factor $\beta$ and harmonic weights $A_0, A_1, A_2$. As could be expected, the truncated solution performs very well for moderate initial conditions $u_0 - 1 \lesssim 0.1$ near the circular orbit but becomes increasingly less accurate for more eccentric orbits where higher harmonics come into play. For example, for $\alpha = 1.5$ (\ref{eqs2harmonics}) produces\\[-2mm]
\par\noindent
\hspace{0.5em}$u_0 = 1.1$ :
\begin{eqnarray}
& & \quad \beta = 1.2244 \,\,(-0.08\%) \quad A_0 = 1.0012 \,\,(-0.0009\%) \nonumber \\
& & \quad A_1 = 0.09918 \,\,(+0.01\%) \quad A_2 = -0.000409 \,\,(-0.7\%) \nonumber
\end{eqnarray}
\par\noindent
\hspace{0.5em}$u_0 = 1.2$ :
\begin{eqnarray}
& & \quad \beta = 1.2235 \,\,(-0.3\%) \quad A_0 = 1.0048 \,\,(-0.01\%) \nonumber \\
& & \quad A_1 = 0.1968 \,\,(+0.06\%) \quad A_2 = -0.0016 \,\,(-2.4\%) \nonumber
\end{eqnarray}
\par\noindent
\hspace{0.5em}$u_0 = 1.5$ :
\begin{eqnarray}
& & \quad \beta = 1.2179 \,\,(-1.8\%) \quad A_0 = 1.027 \,\,(-0.3\%) \nonumber \\
& & \quad A_1 = 0.482 \,\,(+0.7\%) \quad A_2 = -0.0091 \,\,(-13\%) \nonumber
\end{eqnarray} 
where the percentages between brackets indicate the relative deviations from the accurate solution obtained by numerical integration.
\end{document}